\documentclass{PoS}

\RequirePackage{natbib}   

\newcommand\aj{{\it AJ}}
\newcommand\apj{{\it ApJ}}
\newcommand\apjl{{\it ApJ}}
\newcommand\aap{{\it A\&A}}
\newcommand\mnras{{\it MNRAS}}
\newcommand\nat{{\it Nature}}

\newcommand{\e}{\epsilon}
\newcommand{\g}{\gamma}

\newcommand{\ep}{\epsilon^\prime}

\newcommand{\dD}{\delta_{\rm D}}

\newcommand{\psim}{\lower.5ex\hbox{$\; \buildrel \propto \over\sim \;$}}
\newcommand{\lbar}{\lower.0ex\hbox{$\; \buildrel {\lower0.0ex \hbox{-}} \over\lambda  \;$}}

\title{Black-Hole Engine Kinematics, Flares from PKS 2155-304, and 
Multiwavelength Blazar Analysis}

\ShortTitle{Engine Kinematics and PKS 2155-304}

\author{\speaker{Charles D.\ Dermer} and 
        Justin D. Finke \thanks{NRL/NRC Research Associate} \\
        U.S. Naval Research Laboratory, Code 7653, 4555 Overlook SW, 
	Washington, DC 20375-5352\\
        E-mail: \email{charles.dermer@nrl.navy.mil}, 
	\email{justin.finke@nrl.navy.mil}}

\author{Govind Menon \\
        Department of Mathematics and Physics, Troy University, Troy, AL, 36082 \\
        E-mail: \email{gmenon@troy.edu}}

\abstract{
 Kinematical and luminosity relations for black-hole jet sources are
reviewed. If the TeV flares observed from PKS 2155-304 in 2006 July
are assumed to originate from a black hole with mass $\approx 10^8 M_8
M_\odot$, then the $\sim 5$ minute variability timescale is consistent
with the light-travel time across the Schwarzschild radius of the
black hole if $M_8\sim 1$.  The absolute jet power in a
synchrotron/SSC model exceeds, however, the Eddington luminosity for a
black hole with $M_8\sim 1$ unless the jet is highly efficient.  The
maximum Blandford-Znajek power is $\sim 10^{46}M_8$ ergs s$^{-1}$ if
the magnetic-field energy density threading the horizon is equated
with the luminous energy density in the vicinity of the black hole.
An external Compton component can relax power requirements, so a black
hole with mass $\sim 10^8 M_\odot$ could explain the observed flaring
behavior.  For the Swift and HESS data taken in 2006 July,
relativistic outflows with bulk Lorentz factor $\Gamma \gtrsim 30$
satisfy $\gamma$-$\gamma$ attenuation limits.  If this system harbors
a binary black hole, then the accretion disk from a more massive,
$\sim 10^9 M_\odot$ black-hole primary would make an additional
external radiation component.  Dual thermal accretion disk signatures
would confirm this scenario.
}

\FullConference{Workshop on Blazar Variability across the Electromagnetic Spectrum\\
		 April 22-25 2008\\
		 Palaiseau, France}

\begin{document}

\section{Introduction}

HESS observations \citep{aha07} of extraordinary $\gamma$-ray activity
in PKS 2155-304 show flaring time scales as short as $t_v \approx 300$
s in a succession of $\sim 200$ GeV -- 4 TeV outbursts lasting more
than 60 minutes on 27 -- 28 July, 2006.  The apparent isotropic TeV
power of the observed radiation during peak activity is $\gtrsim
3\times 10^{46}$ ergs s$^{-1}$. At a redshift $z = 0.116$ and
luminosity distance $d_L \cong 540$ Mpc, attenuation by the
extragalactic background light (EBL) means that the isotropic
$\gamma$-ray power, including GeV $\gamma$-rays, can be a factor $\sim
10$ greater. The TeV spectrum is described by a broken power law with
number indices $\sim 2.7$ -- 3.

RXTE and Swift observations began on July 29th, revealing a
synchrotron component peaking at UV/soft X-ray energies with energy
flux $\sim 5$ -- 10 less than the TeV energy flux \citep{fos07}.  The
optical emission is poorly correlated with the X-ray emission, and the
optical flux can remain high when the X-ray flux becomes low and the
spectrum steep \citep{fos08}.  The $\gamma$-ray energy flux is
comparable to the synchrotron (radio to 10 keV) energy flux in low
states. Estimates of internal $\gamma$-$\gamma$ absorption already
indicate $\dD \gtrsim 50$ \citep{bfr08}.  Detailed synchrotron/SSC
blazar modeling \citep{fdb08} of the RXTE/Swift and HESS
contemporaneous (but not simultaneous) data, taking into account
internal source and $\gamma$-$\gamma$ absorption for different EBL
models, requires large Doppler factors, $\dD \gtrsim 100$, for $t_{v}
= 300$ s.  These results are in accord with large Doppler factors
encountered previously when using a synchrotron/SSC scenario to model
BL Lac objects \citep[e.g.,][]{kca02}.  If the mass of the
supermassive black-hole engine is $\sim 10^8 M_8 M_\odot$, then its
Schwarzschild timescale is
\begin{equation}
t_{\rm S} = {R_{\rm S}\over c} = {2GM\over c^3} = 
990 M_8 \;{\rm s}\;\cong 10^3 M_8\;{\rm s}.
\label{tS}
\end{equation}
Interpreting the variability time scale $t_v = 300 t_{5m}$ s, 
$t_{5m}\sim 1$, as a limit on 
the engine size scale means that the black hole powering the 
TeV emission from PKS 2155-304 has
mass $M_8\sim 0.5$.

Different than the approach of \citet{bfr08}, 
who postulate patchy flaring regions or
enhanced localized emission regions from jet instabilities, or \citet{lev07}, 
who considers jet deceleration by radiative drag from 
localized radiation fields near the jet, we consider implications 
of requiring that the engine
size scale, characterized by the $R_{\rm S}$, 
be defined in terms of the measured 
variability time scale $t_v$.

In Section 2, the kinematics of relativistic outflows from
black-hole jets are reviewed and applied to the flares of 
PKS 2155-304. The implications of the model are described in 
Section 3, and summary and conclusions are given in Section 4.

\section{Kinematics}

The causality condition for variability is that for 
large amplitude flaring, light travel time limitations
restrict
the temporal variability to timescale $\Delta t^\prime
\gtrsim \Delta r^\prime/c$ 
in the comoving frame. 
From the time transformation, 
where $z$ is the source redshift, $t_v = (1+z)\Delta t^\prime/\dD$, so 
\begin{equation}
t_v \gtrsim \frac{\Delta r^\prime}{c\delta_D} (1+z)\;\; \rm{or}\;\; 
\Delta r^\prime \lesssim \frac{c\delta_D t_v}{1+z}\; ,
\end{equation}
where $z$ is the source redshift.
In the comoving frame, the characteristic
size scale, $\Delta r^\prime$, must be larger by a factor of $\Gamma$ larger
than stationary frame size scale, $\Delta r_*$, so 
that length contraction in the stationary frame recovers
a size scale given by $\Delta r_*$; thus 
$\Delta r^\prime \cong \Gamma \Delta r_*$. 
Taking the black-hole engine sizescale
\begin{equation}
\Delta r_* \cong r_S\; 
\end{equation}
gives
\begin{equation}
t_v \gtrsim {(1+z)\over c}{\Gamma\over \dD } r_{\rm S}\gtrsim 
{(1+z)\over 2c}\;r_{\rm S}\ .
\label{tv}
\end{equation}
Thus the 5 minute variability time scale of PKS 2155-304 essentially
corresponds to the Schwarzschild radius of a black hole with $\approx
10^8 M_\odot$ \citep[see also][]{lev08}.  

This is in contrast to the reasoning given by \citet{aha07} and the
\citet{alb08}, who take the stationary frame sizescale equal to the
comoving sizescale of the emission region, $\Delta r^\prime \cong
\Delta r_*$.  This gives a lower limit to the Doppler factor but 
is based upon a supposition that the comoving emission sizescale 
corresponds to the Schwarzschild radius of the black hole.

The $\nu F_\nu$ spectral flux from jetted outflow is 
\begin{equation}
f_\e = {\dD^4\over 4\pi d_L^2}\ep L^\prime(\ep )
= {\dD^4\over d_L^2}V_b^\prime \ep j^\prime(\ep,\Omega^\prime )\;,\;
\label{fe}
\end{equation}
so that the comoving luminosity 
\begin{equation}
L^\prime = {4\pi d_L^2 \Phi_E\over \dD^4}\;,
\label{Lprime}
\end{equation}
where $\Phi_E = \int_0^\infty d\e\;\e^{-1}f_\e$ is the measured bolometric energy flux.
The apparent isotropic luminosity is therefore
\begin{equation}
L_{iso} = 4\pi d_L^2 \Phi_E = 4\pi d_L^2\int_0^\infty d\e\;{f_\e\over \e}\;.
\label{Liso}
\end{equation}

The absolute luminosity in electromagnetic radiation 
is related to the measured apparent isotropic luminosity
through the beaming factor $f_b$, so that $L_{abs} = f_b L_{iso}$.
The beaming factor for a two-sided top-hat jet source is 
\begin{equation}
f_b \cong 1-\cos\theta_j \stackrel{\Gamma \gg 1}
{\;\;\;\rightarrow\;\;\;\;} 
 {\theta_j^2\over 2} \cong {1\over 2\Gamma^2}\;.
\label{fb}
\end{equation}
Thus $f_b^{-1} \cong 2\Gamma^2 \cong 200,~1800$ for $\Gamma = 10,~30$.
The analysis of \citet{jor05} for radio galaxies, BL Lac objects,
and flat spectrum quasars gives $\theta_j \cong 0.6/\Gamma$. Note also 
that \citet{nie08} use an expression corresponding to eq.\ (\ref{Lprime})
to relate the measured energy flux to the absolute jet luminosity, though
this actually relates it to the comoving luminosity.

These luminosities can be compared to the Eddington luminosity, 
assuming that the jet power ultimately derives from the accretion 
power. The Eddington luminosity
\begin{equation}
L_{\rm Edd} = {4\pi G M m_H c\over \sigma_{\rm T}}
\cong 1.26\times 10^{46} M_8 \;\;{\rm ergs~s}^{-1}\;.
\label{ledd}
\end{equation}
The Eddington ratio
\begin{equation}
\ell_{\rm Edd} = {\eta_f\dot m c^2\over L_{\rm Edd}}\;,
\label{ellEdd}
\end{equation}
where $\eta_f$ is the efficiency to transform the gravitational 
potential energy of accreting matter to
escaping radiation. 

If jets are powered by accretion, assumed to 
be Eddington-limited, then $L_{iso} \cong L_{abs}/f_b 
\ll 2\Gamma^2 L_{\rm Edd}$, so 
\begin{equation}
\Gamma \gg \sqrt{{4\pi d_L^2 \Phi_E\over 2L_{\rm Edd}}}
\cong 1.2 \sqrt{{\Phi_{-9}}
\over M_8}
\cong 1.5 \sqrt{{\Phi_{-9}}
\over t_{v}(300{\rm~s})}
\label{Gamma}
\end{equation}
and $\Phi_{-9}$ is the bolometric source $\gamma$-ray luminosity
in units of $10^{-9}{\rm~ ergs~cm}^{-2}{\rm~s}^{-1}$.

Corrections for the EBL and the addition of the GeV energy flux,
the level of 
which will be determined from GLAST observations,
could show that $\Phi_{-9} \sim
10$, so that $\Gamma\gtrsim 5$. Already from $\gamma$-$\gamma$ arguments,
however, $\Gamma\gtrsim 30$.  For example, the minimum Doppler factor 
from the requirement that the emission region is 
transparent, namely $\tau_{\g\g}(E) < 1$, 
is given for a broken power-law target synchrotron photon spectrum 
by the expression
\begin{equation}
\dD \gtrsim 
\cases{29{\e_6^{0.23} t_{5m}^{-0.14}}\;, &  $\e_6 < t_{5m}^{-1/2}$
 $~$ \cr\cr 
 29.5{\e_6^{0.15} t_{5m}^{-0.18}}\;,\; & $\e_6 > t_{5m}^{-1/2}$
 \cr}\;
\label{rLarmor}
\end{equation} 
  \citep[e.g.,][]{dg95,drl07}, where $10^6\e_6m_ec^2$ is the photon energy, so that $\e_6=1$ corresponds to 
$\approx 500$ GeV photons. When $\e_6\approx 10$, $\dD \gtrsim 50$, 
in agreement with \citet{bfr08}  and \citet{fdb08},
though the multi-TeV emission is not certain to be as variable as the more numerous
photons near the low-energy threshold of the $\gamma$-ray 
telescopes unless one assumes that the variability is attributed to cooling.

\begin{figure}[t]
\center
{\includegraphics[scale=0.5]{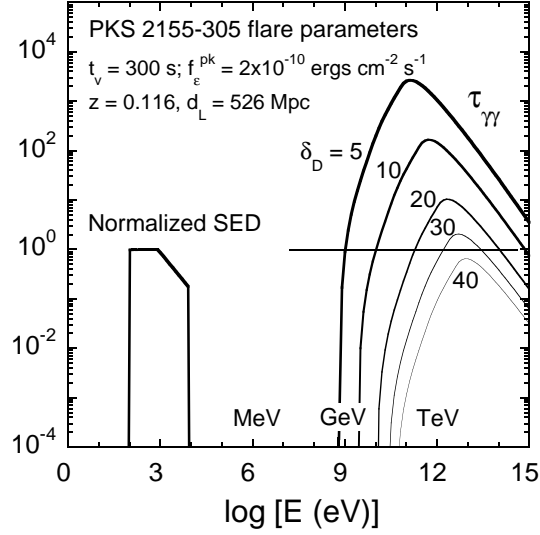}}
\caption{Normalized SED for the target photons of PKS 2155-304, and 
the optical depth $\tau_{\gamma\gamma}$ to absorption of $\gamma$ rays with 
energy E off the target photon distribution as a function of the Doppler 
factor of the outflow.}  
\label{fig1}
\end{figure} 

Fig.\ 1 shows a calculation of $\tau_{\gamma\gamma}$ using a target photon synchrotron
spectrum approximated by the normalized SED shown in the figure (compare Fig.\ 2)
A broader synchrotron spectrum would more effectively attenuate the $\gamma$ rays, 
but even with this minimal target SED, $\dD \gtrsim 30$ is required for $\tau_{\g\g} < 1$ 
for photons observed with 1 TeV energies.

\section{Model Implications}

Already we have assumed a number of specific models: a top-hat jet for
the beaming factor, and a standard jet in the $\gamma$-$\gamma$
constraint on $\dD$, where electrons and photons are isotropically
distributed in a plasma with random magnetic field directions in a
fluid well defined by a single bulk flow Lorentz factor.  Our starting
hypothesis is that the engine timescale sets the variability
timescale.  We furthermore assume that lineless BL Lac objects like
PKS 2155-304, Mrk 421, and Mrk 501 (unlike BL Lac itself) are well
described by a synchrotron/SSC model.

We have recently completed a detailed synchrotron/SSC 
spectral analysis  
of PKS 2155-304 and Mrk 421 \citep{fdb08}. Our approach
differs from the standard technique of injecting electrons
and evolving them in response to energy losses. Instead we
use the lower-energy synchrotron portion of the spectrum 
to infer the electron distribution function, and use this
distribution to calculate the SSC radiation. 
The best-fit model is obtained from  a numerical $\chi^2$ minimization technique, 
given a specified intensity of the (in general, $z$-dependent) EBL.
With this approach, we can precisely calculate the absolute jet power and 
the source luminosity and obtain the radiative efficiency.

Our solutions have a very small radiative 
efficiency; in order to fit both synchrotron optical and X-rays
and TeV $\gamma$ rays in a synchrotron/SSC model for PKS 2155-304,
we find that
the magnetic field is required to be nearly two orders of magnitude
below the equipartition strength. An example of the 
spectral fitting is shown in Fig.\ 2.
Absolute jet powers $\gtrsim 3\times 10^{46}$
ergs s$^{-1}$ are required in the dashed curve in Fig.\ 1 
\citep[Model 8 of][]{fdb08}
with $t_{5m} = 1$, the EBL of \citet{pbs05}, and $\gamma_{min} = 1000$, 
where $\gamma_{min}$ is the minimum electron
Lorentz factor of the electron distribution.
The dashed curve shows the same model except with 
$\gamma_{min} = 5000$.  The higher low-energy cutoff leads to 
a change for absolute jet power from $P_j = 2.8\times 10^{46}$ ergs s$^{-1}$
to $P_j = 8.1\times 10^{45}$ ergs s$^{-1}$. The deduced Doppler factor is
$\dD = 107$ in both cases.

\begin{figure}[t]
\center
{\includegraphics[scale=0.3]{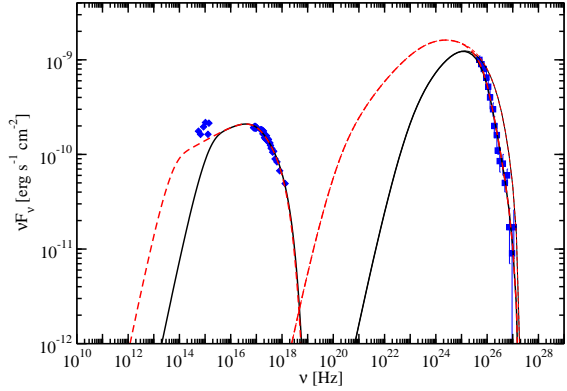}}
\caption[]{Spectral model for PKS 2155-304 for different parameters.
Dashed curve: Model 8 from \citet{fdb08}. Solid curve:
Same as Model 8 but with $\gamma_{min} = 5000$ rather than 
$\gamma_{min} = 1000$.}  
\label{fig2}
\end{figure} 

Jet powers of this magnitude seems to violate our assumption that the
accretion luminosity is Eddington-limited, because it requires a mass
$M_8\gtrsim 1$ to power the jet with, moreover, $\approx 100$\% efficiency. 
Choosing a greater mass would be even more inconsistent with the measured variability
timescale, which by hypothesis sets the engine size scale and the
black hole mass.  As we have seen, some reduction in the jet power can be achieved by
fitting only the X-rays and significantly under-fitting the
lower-frequency optical/UV synchrotron emission \citep[see also][]{gt08}.

The unrealistically high efficiency means either that the model breaks
down or that the basic assumption of sub-Eddington luminosity is
wrong.  Eddington ratios much larger than unity may be generated by
rapid flares representing enhanced accretion episodes, for example, an
accretion instability or the capture of a star. In GRBs this ratio is
about $10^{14}$, but the accretion torus thought to power GRBs
represents a very different system than the gas accreting on the
supermassive black hole of an AGN. The spectrum and intensity of the
accretion-disk emission could reveal the accretion luminosity, but is
difficult to detect from a blazar during flaring periods. During low
states, or for misaligned jets like 3C 273, the accretion disk
radiation can be detected.  For example, during a low state of 3C 279,
a flat spectrum radio quasar, \citet{pia99} detected the disk emission
at the level of $\sim 10^{46}$ ergs s$^{-1}$ in UV.  Detection of disk
emission from BL Lac objects has proven more difficult.

The conflict between the masses inferred from 
the variability timescale 
and the jet power 
is, unfortunately, not resolved if the assumption that accretion 
is the source of the jet power is wrong. In the spin paradigm \citep{bla90,dm09},
black-hole jets are powered by the rotational energy stored in 
the spinning black hole. The Blandford-Znajek mechanism extracts
that energy through electrodynamic processes in the black hole 
magnetosphere.
The BZ power 
\begin{equation}
L_{BZ} \cong 10^{45}\varepsilon B_4^2 M_8^2\;{\rm ergs~s}^{-1}\;
\label{LBZ}
\end{equation}
\citep{lev06}, where $B_4$ is the mean magnetic field near the black hole in units
of $10$ kG, and $\varepsilon \lesssim 0.1$ is an efficiency factor.

We reproduce this relation by a qualitative estimate. The 
 equipartition magnetic
field and related Poynting
power can be estimated by 
treating the black hole magnetospheric system as a 
magnetic dipole produced by a current
flowing near the black hole (analogous to a pulsar model). 
 The magnetized accretion disk of the black hole rotates at the rate
$\Omega(R_{0})$ at radius $R_{0}$ from the center of the black hole.
$R_{0}$ may coincide with the radius where energy dissipation is
largest, the innermost radius defined by the spin of the black hole,
or the inner edge of the accretion disk.  Energy is lost when the
magnetic-field lines are disrupted at the light cylinder of radius
$R_{\rm LC} = c/\Omega_{\rm LC}$, corresponding to a transition from
the near-field to the wave field. For a dipole magnetic field,
$B(R_{LC}) \simeq B_0(R_{0}/ R_{LC})^3$ with $B_0 = B(R_{0})$, so that
the energy-loss rate in Poynting flux is
\begin{equation}
P_{\rm P} \cong 4\pi R_{LC}^2 {B_{0}^2\over 8\pi}\;c\; 
\left( {R_{0}\over R_{LC}}\right)^6 \; \cong \; 
{c\over 2} B_0^2 R_0^6 R_{LC}^{-4}\;,
\label{dedtp}
\end{equation}
where the light-cylinder frequency is identified with the Keplerian
orbital frequency at radius $R_0$; thus $\Omega_{LC} \approx
\Omega_{\rm K}(R_0)= \sqrt{GM/R^3}$, and $\tilde R_{LC} = \tilde
R_0^{3/2}$. (Tildes mean that the distances are
rescaled in units of gravitational radii $R_g = GM/c^2$.)

The magnetic field is defined with respect to the observed radiant
luminosity $L_{rad} = \ell_{\rm Edd}L_{\rm Edd}$ (eq.\ [\ref{ledd}]) and size $R_0$ using
the equipartion magnetic field which we define here in terms
of the luminous energy density, giving
\begin{equation}
{B_0^2\over 8\pi} = {L_{rad}\over 4\pi R_0^2 c} 
\cong {\ell_{\rm Edd}L_{\rm Edd}\over 4\pi R_0^2 c} \;,
\label{equipart3}
\end{equation}
or
\begin{equation}
B_0({\rm Gauss}) \;\cong {6\times 10^4\over \tilde
R_0}\;\sqrt{{\ell_{\rm Edd}}\over M_8}\;.
\label{B0SMBH}
\end{equation}
This suggests that the inner magnetized regions surrounding
supermassive black holes should have ${\cal O}($kG) magnetic fields. Eqs.\
(\ref{dedtp}) and (\ref{equipart3}) together imply
\begin{equation}
P_{\rm P}({\rm ergs~s}^{-1})\; \simeq 10^{46}\; {M_8 \ell_{\rm
Edd}\over \tilde R_0^2}\;.
\label{equipart4}
\end{equation}
For accretion at $\ell_{\rm Edd}\sim 1$-10\% and energy dissipation at
$\tilde R_0 \sim 6$ (Schwarzschild metric) and $\tilde R_0 \sim 3$ (rotating
black hole), the Poynting power as estimated here is
unlikely to be capable of satisfying jet power requirements
in a synchrotron/SSC model for supermassive black-hole jets.

A more detailed derivation of the Blandford-Znajek power can be
obtained by solving the constraint equation for black-hole
electrodynamics \citep{md05,dm09}, written in the 3+1 formalism of
\citep{kom04}.  There a solution for the fields and currents was found
that generalizes the Blandford-Znajek split monopole solution to
accommodate the case of a black hole for all values of $a^2 < M^2$,
where $a$ is the spin paramter \cite{md05}.  The rate of energy
extraction from the black-hole spin for this $\Omega_+$ solution is

\begin{equation}
P_{\Omega_+} = \frac{\pi Q_0^2}{a r_+}\left[ \arctan \frac{a}{r_+}- \frac{a}{2M}\right]\;.
\label{fracdEdt}
\end{equation}

The charge $Q_0$ can be related to the 
magnetic field intensity $\bar B$ 
threading the event horizon of the
black hole at the equator. This leads to 
a maximum Blandford-Znajek power 
\begin{equation}
P_{\Omega_+} \;\cong \;2\times 10^{46}\; \ell_{Edd} M_8\; \;{\rm~ergs~s}^{-1}\;
\label{fracdEdt5}
\end{equation}
for the 
$\Omega_+$ solution. 
Thus we see that the Blandford-Znajek process cannot extract energy at
a rate much larger than the Eddington luminosity for a system whose radiative
output is limited by the Eddington luminosity.

Rather than reject the hypothesis that the engine size scale determines the 
minimum variability time scale, we can in principle 
resolve the contradiction by abandoning the simple one-zone synchrotron/SSC
model. Radiative efficiency in 
external Compton models improve with larger $\Gamma$ values, 
as noted by \citet{bfr08} and \citet{gt08},  
but the target radiation field provides additional opacity
that must be carefully considered \citep{dfkb08}.
Other soft photon sources, for example, 
synchrotron radiation from the extended 
decelerating jet \citep{gk03}, or radiation from a slower moving
sheath around a faster moving spine \citep{tg08}
could, if good spectral fits in such models can 
be obtained, relax the jet power constraints.
Detailed spectral analyses and modeling studies
are needed to determine if power requirements can be 
reduced in a jet blazar powered by $\sim 10^8$ Solar mass
black hole by adding an external target radiation 
field. 

\section{Summary and Speculations}

Until 
simultaneous optical, Swift, GLAST, and HESS data are available 
to model, it is uncertain whether an $\sim 10^8 M_\odot$ black hole 
can explain the variability and 
spectral behavior of PKS 2155-304. Supposing that it can, one question of interest 
is then how to 
satisfy the bulge/black-hole mass 
relation for the mass of the black hole in PKS 2155-304, 
which is estimated to be $\approx 1$ -- $2 \times 10^9 M_\odot$ \citep{aha07,kot98}.
The dispersion in the black-hole/bulge luminosity relationship 
for AGN indicates that only $\lesssim 5$\% of the sources 
are more than one order of magnitude away from the
best-fit line \citep{md02}. For PKS 2155-304 not to be an outlier on 
this plot, we assume that the mass of the black-hole system of this source exceeds
$10^9 M_\odot$.

This can be realized if the $10^8 M_\odot$ black hole is a
member of a binary system containing a larger $\sim 10^9 M_\odot$
black hole. The jetted black hole with its accretion disk may be assumed
to define a fairly specific direction into which plasma is
ejected. The orbital motion in the binary system produces a
Parker-like spiral pattern from the outflowing jets, which may be
related to the spiral rotation of the linear polarization of the
position angle observed in sources like BL Lac \citep{mar08} and OJ
287 \citep[e.g.,][]{vetal98}.  Depending on how the jet precesses
during its orbital motion, a converging collimation shock could be
formed \citep{mar08}.

A binary system of supermassive black holes would display dual accretion disk
signatures, each of which can provide target photons for jet Compton scattering.
As in the case of models of microquasars \citep[reviewed recently by][]{br08}, the external 
Compton $\gamma$-ray 
emission would display periodic signatures due to $\gamma\gamma$ attenuation
and anisotropic Compton scattering. For systems with orbital periods $\lesssim 
 1$ -- 2 years, GLAST can search for variations in spectra with orbital phase.

\section{Acknowledgements}
We thank Berrie Giebels for participation in this workshop,  
and Amir Levinson for useful correspondence and comments about this paper.
The work of CDD is supported by the Office of Naval Research, and
the research of JDF is supported by the GLAST Interdisciplinary Program.

\end{document}